\documentstyle[psfig]{mn}		

\title[2dFGRS: spectral types and luminosity functions]
{The 2dF Galaxy Redshift Survey: Spectral Types and Luminosity Functions}

\author[Folkes et al.]{
\parbox[t]{\textwidth}{Simon Folkes$^1$, Shai Ronen$^1$, Ian Price$^2$,
Ofer Lahav$^{1,10}$, Matthew Colless$^2$, Steve Maddox$^1$, 
Kathryn Deeley$^3$, Karl Glazebrook$^3$, 
Joss Bland-Hawthorn$^3$, Russell Cannon$^3$, Shaun Cole$^4$, Chris
Collins$^5$, Warrick Couch$^6$, Simon P.\ Driver$^{11}$, Gavin Dalton$^7$,
George Efstathiou$^1$, Richard S.\ Ellis$^1$, Carlos S.\ Frenk$^4$, 
Nick Kaiser$^8$, Ian Lewis$^3$, Stuart Lumsden$^3$, John
Peacock$^9$, Bruce A. Peterson$^2$, Will Sutherland$^7$, Keith Taylor$^3$}
\vspace*{6pt} \\ 
$^1$Institute of Astronomy, University of Cambridge, Madingley Road,
Cambridge CB3 0HA, United Kingdom \\
$^2$Research School of Astronomy \& Astrophysics, The Australian 
National University, Weston Creek, ACT 2611, Australia \\
$^3$Anglo-Australian Observatory, P.O.\ Box 296, Epping, NSW 2121,
Australia\\ 
$^4$Department of Physics, South Road, Durham DH1 3LE, United Kingdom \\
$^5$Astrophysics Research Institute, Liverpool John Moores University,  
Twelve Quays House, Egerton Wharf, Birkenhead, L14 1LD, United Kingdom\\
$^6$Department of Astrophysics, University of New South Wales, Sydney,
NSW 2052, Australia\\
$^7$Department of Physics, Keble Road, Oxford OX3RH, United Kingdom\\
$^8$Institute for Astronomy, University of Hawaii, 2680 Woodlawn Drive,
Honolulu, HI 96822, U.S.A.\\
$^9$Institute of Astronomy, University of Edinburgh, Royal Observatory,
Edinburgh EH9 3HJ, United Kingdom\\
$^{10}$Racah Institute of Physics, The Hebrew University, Jerusalem 91904,
Israel\\
$^{11}$School of Physics and Astronomy, North Haugh, St Andrews, Fife, KY6 9SS,
United Kingdom}
\date{Accepted ---. Received ---; in original form ---.}

\newlength{\plotwidth}
\newcommand{\etal}{\hbox{et~al.}}

\newcommand{\eg}{\hbox{e.g.}}

\newcommand{\Mpc}{\hbox{\,h$^{-1}$\,Mpc}}

\newcommand{\sqdeg}{\hbox{\,$\Box^\circ$}}

\newcommand{\gs}
           {\mathrel{\hbox{\rlap{\hbox{\lower4pt\hbox{$\sim$}}}\hbox{$>$}}}}
\newcommand{\ls}
           {\mathrel{\hbox{\rlap{\hbox{\lower4pt\hbox{$\sim$}}}\hbox{$<$}}}}

\begin{document}

\maketitle

\begin{abstract}
We describe the 2dF Galaxy Redshift Survey (2dFGRS), and the
current status of the observations. In this exploratory paper, 
we apply a Principal Component
Analysis to a preliminary sample of 5869 galaxy spectra and use
the two most significant components to split the sample into five
spectral classes. These classes are defined by considering visual
classifications of a subset of the 2dF spectra, and also by comparing
to high quality spectra of local galaxies. 
We calculate a luminosity function for each of the different classes and
find that later-type galaxies have a fainter characteristic magnitude,
and a steeper faint-end slope. For the whole sample we find $M^\star=-19.7$
(for $\Omega$=1,$H_0$=100\,km\,s$^{-1}$\,Mpc$^{-1}$), 
$\alpha=-1.3$, $\phi^\star=0.017$. For class 1 (`early-type') we find $M^\star=-19.6$,
$\alpha=-0.7$, while for class 5 (`late-type') we find $M^\star=-19.0$, 
$\alpha=-1.7$. The derived 2dF luminosity functions agree well with other
recent luminosity function estimates.

\end{abstract}


\section{INTRODUCTION}
\label{sec:intro}

The 2dF Galaxy Redshift Survey (2dFGRS; Colless 1998, Maddox 1998) is a
major new redshift survey utilising the full capabilities of the 2dF
multi-fibre spectrograph on the Anglo-Australian Telescope (AAT). The
observational goal of the survey is to obtain high quality spectra and
redshifts for 250,000 galaxies to an extinction-corrected limit of
$b_J$=19.45. The survey will eventually cover approximately 2000\sqdeg,
made up of two continuous declination strips plus 100 random
2\degr-diameter fields. One strip is in the southern Galactic hemisphere
and covers approximately 75\degr$\times$15\degr\ centred close to the
South Galactic Pole at ($\alpha$,$\delta$)=($01^h$,$-30$\degr); the other
strip is in the northern Galactic hemisphere and covers
75\degr$\times$7.5\degr\ centred at
($\alpha$,$\delta$)=($12.5^h$,$+00$\degr). The 100 random fields are
spread uniformly over a 7000\sqdeg\ region in the southern Galactic cap.

The survey has been designed to provide a detailed picture of the
large-scale structure of the galaxy distribution in order to understand
structure formation and evolution and to address cosmological issues
such as the nature of the dark matter, the mean mass density of the
universe, the role of bias in galaxy formation and the Gaussianity of
the initial mass distribution. However the survey will also yield a
comprehensive database for investigating the properties of the
low-redshift galaxy population, providing: $b_J$ and $r_F$ magnitudes and
various image parameters from the blue and red Southern Sky Survey
plates scanned with the Automatic Plate Measuring machine (APM); 
moderate quality spectra (S/N$\gs$10 per $4.3\AA$ pixel); derived
spectroscopic quantities such as redshifts and spectral types.

The first test observations for the 2dFGRS were taken at the start of
the 2dF instrument commissioning period in November 1996. The first
survey observations with all 400 fibres were obtained in October 1997,
and as of March 1999 we have measured over 40,000 redshifts. We plan to
complete the survey observations by the end of 2000.

This paper presents some of the first results from the 2dFGRS. It deals
with the Principal Component Analysis (PCA) methods that we are
developing for classifying galaxy spectra, and the application of these
methods to deriving the luminosity functions for different galaxy
spectral types.
The luminosity function (LF) is a fundamental characterization of the
galaxy population. It has been measured from many galaxy surveys with
differing sample selections, covering a wide range of redshifts.
Generally a Schechter function (Schechter 1976),
\begin{equation}
\phi(L) dL=\phi^\star ({L \over L^\star})^\alpha exp(-{L \over L^\star}) {dL
\over L^*} ,
\end{equation}
with 
\begin{equation}
{L \over L^\star} = 10^{0.4(M^\star-M)}
\end{equation}
provides a good fit to field galaxy data, 
but the parameters $M^\star$, $\alpha$ and 
$\phi^*$ are still relatively uncertain.

Surveys of bright field galaxies, such as the CfA2 (Marzke et al 1994)
and SSRS2 (Marzke \& Da Costa 1997) have mean redshift $\bar z \sim
0.02$ and so cover relatively small volumes. This leads to a large
sample variance on these measurements.  Also they are based on
photometric catalogues which have been visually selected from
photographic data, which makes selection biases hard to quantify.
Deeper surveys such as the CFRS (Lilly et al 1995), and AUTOFIB
(Ellis et al 1996) surveys are based on better controlled galaxy
samples, but the sample volumes are very small, and they reach to much
higher redshifts ($z\sim1$), so that galaxy evolution is an important
factor. 

The Stromlo-APM Redshift Survey (SAPM, Loveday et al 1992) the 
Las Campanas Redshift
Survey (LCRS; Lin et al 1996) and the ESO Slice Project (ESP; Zucca et
al 1997) cover intermediate redshifts, $z\sim0.05 - 0.2$, and give the
most reliable current estimates of the local luminosity
function. The estimated values of $M^*$ from these surveys agree to
about 0.1 magnitude, and $\phi^*$ varies by a factor of $\sim$ 1.5.
The faint-end slope, $\alpha$, is more poorly determined, ranging from
$-0.7\pm0.05$ for the LCRS to $-1.22^{+0.06}_{-0.07} $ for the
ESP. Some of these differences are likely to be due to sample
variance, but some may be due to different selection effects in the
surveys, e.g, the high surface brightness cut imposed on the LCRS survey.  
The 2dF survey reaches to a similar depth, will cover a
much larger volume than the LCRS survey, and 
so have smaller sample variance, allowing us
to estimate luminosity functions for several sub-samples of galaxies.
The mean surface brightness isophotal detection limit of the
underlying APM catalogue is $b_j=25$.

It is well established that the galaxy luminosity function depends on
the type of galaxy that is sampled.  Morphologically late-type
galaxies tend to have a fainter $M^*$ (in the B-band) and steeper faint-end slope,
$\alpha$, (Loveday et al. 1992, Marzke et al 1998). Spectroscopically,
selecting galaxies with higher [OII] equivalent width leads to a
similar trend (Ellis et al. 1996, Lin et al 1996), as does the
selection of bluer galaxies (Marzke \& Da Costa 1997, Lin et al 1996).
These results reflect the correspondence between spectral and
morphological properties, with galaxies of late type morphology having
stronger emission lines and bluer continuum than galaxies of early
type morphology. However this correspondence is quite approximate and
a considerable scatter exists. 

In this paper we use a classification scheme based on PCA of the
galaxy spectra 
(e.g. Connolly et al. 1995; 
Folkes, Lahav \& Maddox 1996; 
Sodr\'e \& Cuevas 1997; Galaz \& de Lapparent 1998; 
Bromley et al. 1998; Glazebrook, Offer \& Deeley 1998; 
Ronen,  Arag\'on-Salamanca \&  Lahav 1999).
This provides a representation of
the spectra in two (or more) dimensions that highlights the
differences between individual galaxies. The technique is based on
finding the directions in spectral space in which the galaxies vary
most, and so offers an efficient, quantitative means of
classification.  Bromley et al (1998) have studied the variation of
the LCRS LF with spectral type using classes from a similar PCA
technique. They find that late type galaxies have a
steeper faint-end slope than early type galaxies. They also find a clear
density-morphology relation: over half of their extreme early-type
objects are found in regions of high density, whereas these regions
contain less than a quarter of their extreme late-type objects.  

The plan of the paper is as follows. In \S\ref{sec:data} we briefly
summarise the construction of the survey source catalogue, the
capabilities of the 2dF multi-fibre spectrograph, the observing and
reduction procedures employed and the main properties of the subset of
the data which will be used in the rest of the paper. In \S\ref{sec:pca}
we outline the fundamentals of Principal Component Analysis, describe
the steps required to prepare the spectra, and then present the
principal components for our sample of galaxy spectra and their
distribution. Two methods to connect the PCA decomposition of a galaxy's
spectrum with its physical (spectral or morphological) type are
investigated in \S\ref{sec:class}. We then define spectral types based
on the PCA decomposition and use these types to derive K-corrections for
each galaxy in our sample. The luminosity functions (LFs) for the whole
sample, and for each spectral type separately, are derived in
\S\ref{sec:lfs}, using both a direct estimator and a parametric method.
Our results are discussed in \S\ref{sec:discuss}, focusing on the
strengths and weaknesses of the PCA spectral classifications and a
comparison of the LFs for our spectral types with similar analyses in
the literature.

\section{THE DATA}
\label{sec:data}

\subsection{Source catalogue}
\label{ssec:source}

The source catalogue for the survey is a revised and extended version of
the APM galaxy catalogue (Maddox et~al.\ 1990a,b,c). This catalogue is
based on APM scans of 390 IIIa-J plates from the UK Schmidt Telescope
(UKST) Southern Sky Survey. The magnitude system for the Southern Sky
Survey is defined by the response of Kodak IIIa-J emulsion in
combination with a GG395 filter, zero pointed using Johnson B band CCD photometry. The extended version of the APM catalogue includes over
5 million galaxies down to $b_J$=20.5 in both north and south Galactic
hemispheres over a region of almost 10$^4$\sqdeg\ (bounded approximately
by declination $\delta$$\leq$+3\degr\ and Galactic latitude
$|b|$$\gs$20\degr). The astrometry for the galaxies in the catalogue has
been significantly improved, so that the rms error is now 0.25\arcsec\
for galaxies with $b_J$=17--19.45. Such precision is required in order to
minimise light losses with the 2\arcsec\ diameter fibres of 2dF. The
photometry of the catalogue is calibrated with numerous CCD sequences
and has a precision (random scatter) of approximately 0.2~mag for 
galaxies with
$b_J$=17--19.45. The mean surface brightness isophotal detection limit of the
APM catalogue is $b_j=25$. The star-galaxy separation is as described in Maddox
et~al.\ (1990b), supplemented by visual validation of each galaxy image.
A full description of the source catalogue is given in Maddox \etal\
(1998, in preparation).

\subsection{The 2dF multi-fibre spectrograph}
\label{ssec:2dF}

The 2dF facility consists of a prime focus corrector, a focal plane
tumbler unit and a robotic fibre positioner, all mounted on an AAT
top-end ring which also supports the two spectrographs. 

The 2dF corrector introduces chromatic distortion (i.e, shifts between
the blue and red images). This is a function of radius and is,
by design, a maximum at the halfway 
point (radius=30 arcmin) and minimum at the
centre and edge.

The absolute shifts, at a particular wavelength, are allowed for 
in the 2dF configuration program and for our survey we configure for 5800\AA. 
Relative to this we have maximum residual shifts of +0.9 arcsec (at 4000\AA) and
-0.3 arcsec
(at 8000\AA). The typical shift, averaged over the field, will be of order half
these
values (Bailey and Glazebrook 1998).

The 2dF instrument has two field plates and two full sets of 400 fibres. 
This allows one plate to be configured while observations are taking 
place with the other. The robotic positioner uses a
gripper head which places magnetic buttons containing the fibre ends at
the required positions on the plate. When the plate is prepared and one
set of observations are complete, the tumbler can rotate to begin
observations on the new field. Each field plate has 400 object fibres
and an additional 4 guide-fibre bundles, which are used for field
acquisition and tracking. The object fibres are 140\,$\mu$m in diameter,
corresponding to about 2.16\arcsec\ at the centre and 2\arcsec\ at
the edge of the field. Two identical spectrographs receive 200 fibres
each and spectra are recorded on thinned Tektronix 1024 CCDs.

Further detail on the design and use of the instrument is given in
Bailey \& Glazebrook (1999), Lewis et al. (1998) and Smith \& Lankshear (1998).

In terms of spectral analysis, the design of the system has a number of
implications. Firstly, all 400 spectra taken in a particular observation
have the same integration time, regardless of the brightness of
individual targets. After the fibre feed, each of the two sets of 200
spectra pass through an identical optical system, which ensures some
consistency. The combination of the fibre size, small positioning errors
and chromatic variation means that, particularly for nearby extended
objects, the observed spectrum is not necessarily representative of the
whole object. The impact of some of these effects is examined in more
detail below.

\subsection{Observations and reductions}
\label{ssec:obsvns}

This paper uses only the data taken for the 2dFGRS in two early
observing runs: 1997 October 29 to 1997 November 3 and 1998 January
23--29. In the former we observed 14 2dF fields and 3882  
galaxies; in the latter we observed 16 2dF fields and 4482 
galaxies. Counting repeats only once this gives a total
of 7972 galaxies. The 400 fibres in each field were
shared between the targets of the 2dFGRS and the 2dF QSO Redshift Survey
(Boyle 1998). The observations were performed with the 300B gratings,
giving an observed wavelength range of approximately 3650\AA\ to
8000\AA, although this varies slightly from fibre to fibre. The spectral
scale was 4.3\AA/pixel and the FWHM resolution measured
from arc lines was 1.8-2.5 pixels (varying over the wavelength range). 

The exposure time for the observations described here was around 70 min. This
time was determined by the time necessary for configururation of the fibres
(with improvement in the configuration time, the observation time has been
reduced to around 60 min). The spectra were reduced using an early version of the {\tt 2dfdr}
pipeline software package (Bailey \& Glazebrook 1999). Typically,
three to four sub-exposures were taken and cosmic rays removed by
a sigma-clipping algorithm. Galaxies at the survey limit of $b_J$=19.45 have a
median S/N of $\sim 14$, which is
more than adequate for measuring redshifts and permits reliable spectral
types to be determined, as described below.

Redshifts were found by two independent methods: the first, 
cross-correlating the spectra  with absorption
line templates, and the second, by emission line fitting. 
These automatic redshift estimates were then
confirmed by visual inspection of each spectrum, and the more reliable of the
two results chosen as the final redshift. A quality flag (Q)
was manually assigned to each redshift: Q=3 and Q=4 (7180 objects) correspond
to reliable redshift determinations;
Q=2 (574 objects) means a probable redshift; and Q=1 (218 objects)
means no redshift could be determined. We note that some stars enter 
the sample because the star-galaxy 
classification criteria for the
source catalogue are chosen to exclude as few galaxies as possible, at
the cost of 4\% contamination of the galaxy sample by stars. As a crude
criterion, and for the purpose of this work, we consider all objects
with $z<0.01$ as stars.  Only the
6899 non-stellar objects with quality flag Q=3 or Q=4 are included in
subsequent analysis. Using this criterion, the sample of galaxies with
reliable redshifts is 90\% of all observed objects not identified as stars. 
We note that more recent observations are giving a higher redshift 
completeness. 

It is worth noting that PCA and redshift determination can
be performed simultaneously in a single procedure which can improve the
redshift determination by reducing its dependency on a predetermined set
of template spectra (Glazebrook \etal\ 1998).  We are
investigating future use of this method.

\begin{figure}
\centerline{\psfig{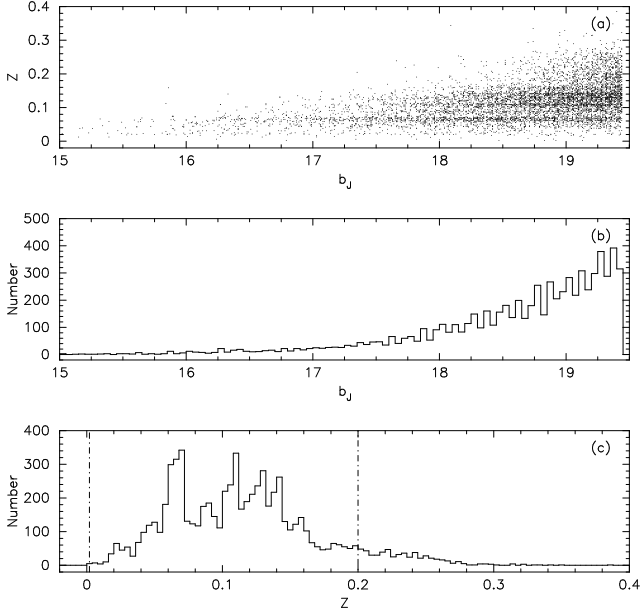}}
\caption{Properties of the sample: (a)~the distribution of redshifts as
a function of apparent magnitude; (b)~the differential number counts;
(c)~the redshift distribution.}
\label{redhist}
\end{figure}

The fields observed in these two runs lie in the northern and southern
declination strips covered by the survey. Note that at the median
redshift of the sample, $\hat z=0.1$, 2\degr\ corresponds to a comoving
distance of 9.7\Mpc. 
Figure~\ref{redhist} summarises some of the properties of this sample.
The distribution in apparent magnitude and redshift is shown in
Figure~\ref{redhist}a; the median redshift of the galaxies increases
from $\hat z=0.07$ at $b_J$=17 to $\hat z=0.14$ at
$b_J$=19.45. The redshift distribution in Figure~\ref{redhist}c
shows considerable clustering in redshift space, reflecting in part 
that our survey is still dominated by a few lines-of-sight whcih intersect
common structures. We expect this to average out when we complete
the full volume.

\begin{figure}
\centerline{\psfig{figure=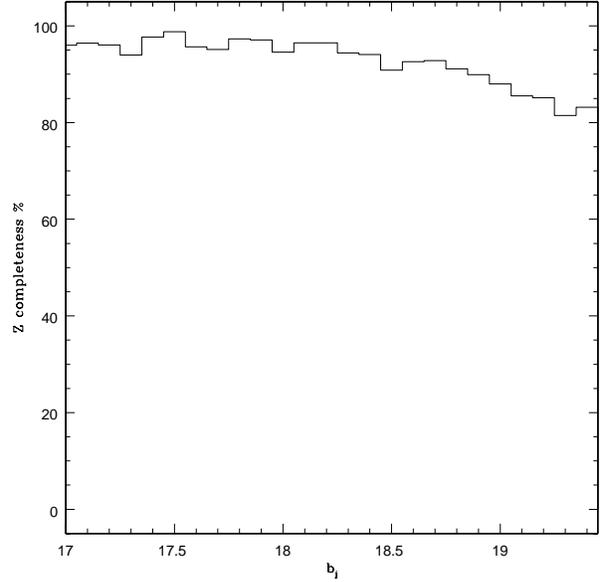,width=\plotwidth}}
\caption{The fraction of objects for which reliable redshifts were obtained
over all observed objects. }
\label{zcomplete}
\end{figure}

Figure~\ref{zcomplete} shows the redshift completeness, as a function
of apparent magnitude, computed as the fraction of objects with good
redshift ($Q>2$) out of all observed objects. 

A further completeness factor must also be included to allow for the
fact that not all galaxies in the photometric catalogues can be
assigned a fibre in the 2dF configurations. One of the constraints on
the configuration is the fact that two fibres cannot be put closer 
than $25''$, for the most favourable geometry. The tiling process of
the fields 
is designed to maximise the completenes of the survey as galaxies
in close pairs can be observed in diffrent tiles. The
final completeness is extremely high (93\%). 

\section{PRINCIPAL COMPONENT ANALYSIS}
\label{sec:pca}

\subsection{The method}
\label{ssec:method}

A spectrum, like any other vector, can be thought of as 
a point in an $M$-dimensional parameter space. One may
wish for a more compact description of the data.
This can be accomplished by Principal
Component Analysis (PCA), a well-known statistical tool that has been
used in a number of astronomical applications (Murtagh \& Heck 1987).
By identifying the {\em linear}
combination of input parameters with maximum variance, PCA finds the
principal components that can be most effectively used
to characterise the inputs.

The formulation of standard PCA is as follows. Consider a set of $N$
objects ($i=1,N$), each with $M$ parameters ($j=1,M$). If $r_{ij}$ are
the original measurements of these parameters for these objects, then
mean subtracted quantities can be constructed,
\begin{equation}
X_{ij} = r_{ij} - {\bar r_j} \ ,
\label{eqn:xij} 
\end{equation}
where ${\bar r_j} = {1 \over N} \sum_{i=1}^{N} r_{ij}$ is the mean. The
covariance matrix for these quantities is given by
\begin{equation}
C_{jk} = {1 \over N} \sum_{i=1}^{N} X_{ij} X_{ik} \hskip0.5cm 1 \leq j \leq M \hskip0.5cm 1 \leq k \leq M \ .
\end{equation}

It can be shown that the axis (i.e, direction in vector space) 
along which the variance is maximal is
the eigenvector ${\bf e_1}$ of the matrix equation
\begin{equation}
C {\bf e_1} = \lambda_1 {\bf e_1} \ , 
\end{equation}
where $\lambda_1$ is the largest eigenvalue (in fact the variance along
the new axis). The other principal axes and eigenvectors obey similar
equations. It is convenient to sort them in decreasing order (ordering by
variance), and to
quantify the fractional variance by
$\lambda_\alpha/\sum_\alpha\lambda_\alpha$. The matrix of all the
eigenvectors forms a new set of orthogonal axes which are ideally suited
to an efficient description of the data set using a truncated
eigenvector matrix employing only the first $P$ eigenvectors
\begin{equation}
U_P=\{e_{jk}\} \hskip1cm 1 \leq k \leq P \hskip1cm 1 \leq j \leq M \ ,
\end{equation}
where $e_{jk}$ is the $j$th component of the $k$th eigenvector.  The
turncation turns out to be efficient because as it happens the cloud of points
which represent the spectra closely lies in a low dimensional sub-space.
This can be seen from the fact that the first few eigenvalues account for
most of the variation in the data, and it can also be seen that 
the higher eigenvectors contain  mostly the noise (Folkes, Lahav \& Maddox, 
1996).

Now if a specific spectrum is taken from the matrix defined in
Equation~\ref{eqn:xij}, or possibly a spectrum from a different source
which has been similarly mean-subtracted and normalised, it can be
represented by the vector of fluxes ${\bf x}$. The projection vector
${\bf z}$ onto the $M$ principal components can be found from (here ${\bf x}$ and
${\bf z}$ are row vectors):
\begin{equation}
{\bf z} = {\bf x} U_M \ .  
\end{equation}
Multiplying by the inverse, the spectrum is given by
\begin{equation}
{\bf x} = {\bf z} {U_M}^{-1} = {\bf z} {U_M}^t \ , 
\end{equation}
since $U_M$ is an orthogonal matrix by definition. However, using only
$P$ principal components the reconstructed spectrum would be
\begin{equation}
{\bf x}_{rec} = {\bf z} {U_P}^t \ , 
\end{equation}
which is an approximation to the true spectrum.

The eigenvectors into which we project the spectra can be viewed as 
`optimal filters' of the spectra, in analogy with other 
spectral diagnostics such as colour filter or spectral index.
Finally, we note that there is some freedom of choice as to whether 
to represent a spectrum as a vector of fluxes or of photon counts. The
decision will affect the resulting principal components, as a representation 
by fluxes will give more weight to the blue end of a spectrum than 
a representation by photon counts. In this paper all spectra
are represented as photon counts, but we leave open the question of which
representation is `the best' in some sense.

\subsection{Data preparation}
\label{ssec:dataprep}

Before we can carry out Principal Component Analysis a number of
procedures are required to prepare the spectra.

Firstly, residuals from strong sky lines and bad columns were removed by
interpolating the continuum across them. Secondly, we corrected for sky 
absorption
at the A and B bands (7550\AA\ to 7700\AA\
 and 6850\AA\ to 6930\AA\, respectively)
as follows: the spectrum was smoothed with a 150\AA\ Gaussian filter to
give the low-resolution spectral shape, and with a 3\AA\ Gaussian filter
to give a noise-reduced spectrum closely following the shape of the
absorption band profiles. The original spectrum was then multiplied by
the ratio of the former to the latter over the spectral ranges covered
by the atmospheric absorption bands.
 
\begin{figure}
\centerline{\psfig{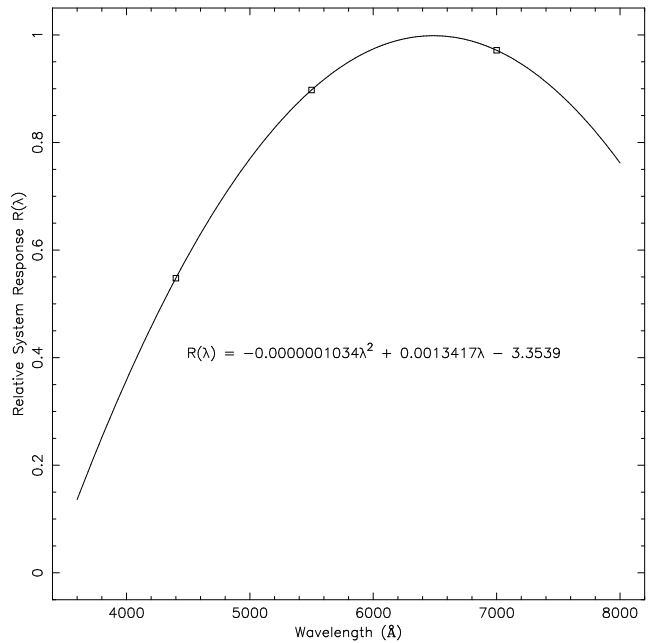}}
\caption{Fitted response function for the 2dF system, based on
observations of Landolt standard stars.}
\label{response}
\end{figure}

The system response also needs to be removed from the spectra. This was done 
by calculating a second-order
polynomial fit to the 2dF system response with the 300B grating from
observations of Landolt standard stars in BVR. 
The fit can be seen in Figure~\ref{response}. Some 
fibre-to-fibre variations in the response function can be expected, as well as
possible time variations. We have observed wavelength dependent variations 
at the level of 20\%. We expect improvement in this with the application of
an improved extraction algorithm (not applied to the data presented here). 
It is possible to remove such unknown variations
in the flux calibration by, e.g., by removing the low frequency 
Fourier components, as was done by Bromley et al. 
(1998). However such a procedure will also remove potentially important
information inherent in the continuum. Therefore, in this paper, we chose to use
our measured flux calibration and retain the whole spectrum. We leave it for a
future study to compare the two approaches.

\begin{figure}
\centerline{\psfig{figure=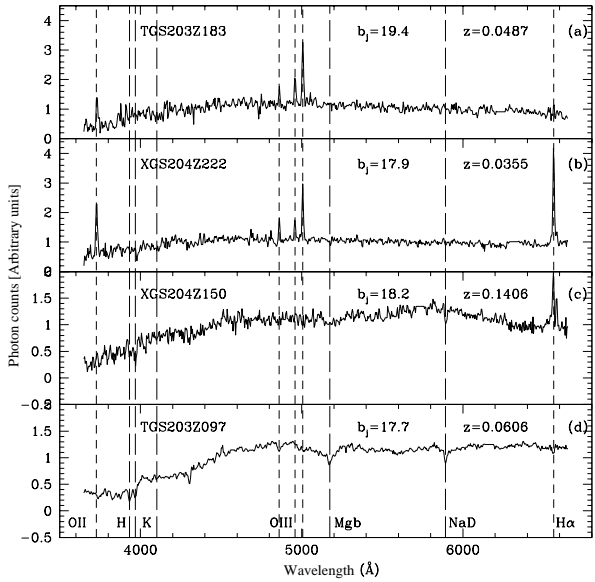,width=\plotwidth}}
\caption{Examples of 2dF spectra prepared for PCA analysis. Sky lines
and bad columns have been interpolated over, atmospheric absorption features
corrected and the instrument response function divided out. The spectra
have then been transformed to the rest-frame, re-sampled to 4\AA\ bins
over the limited range 3700\AA\ to 6650\AA, and normalised to unit mean
flux.}
\label{ready}
\end{figure}

The next step is to de-redshift the spectra to their rest frame and
re-sample them to a uniform spectral scale with 4\AA\ bins. Since the
galaxies cover a range in redshift, the rest-frame spectra cover
different wavelength ranges. To overcome this problem, only the 6015
objects with redshifts in the range 0.01$\le$$z$$\le$0.2 are included
in the analysis. 
All the objects  meeting this criterion
then have rest-frame spectra covering the range 3700\AA\ to 6650\AA\ (the lower limit was  
chosen to exclude the bluest end of the spectrum where the response function is
poor). 
Limiting the analysis to this common wavelength range means that all the
major optical spectral features between [OII] (3727\AA) and H$\alpha$
(6563\AA) are included in the analysis. In order to make the PCA
spectral classifications as robust as possible, objects with redshifts
but relatively low S/N were eliminated by imposing a minimum mean flux
of 50 counts per bin. The spectra are then normalised so that the mean
flux over the whole spectral range is unity. Figure~\ref{ready} shows
examples of the prepared spectra for a range of galaxy types, with some
of the major spectral features indicated. The spectral classifications
and luminosity functions are derived from this final sample of 5869
galaxies, 
each described by  738 spectral bins.

Finally, we reiterate that we applied the PCA analysis to the spectra given as
photon counts per bin, as opposed to energy flux per bin.

\subsection{Application}
\label{ssec:apply}

\begin{figure}
\centerline{\psfig{figure=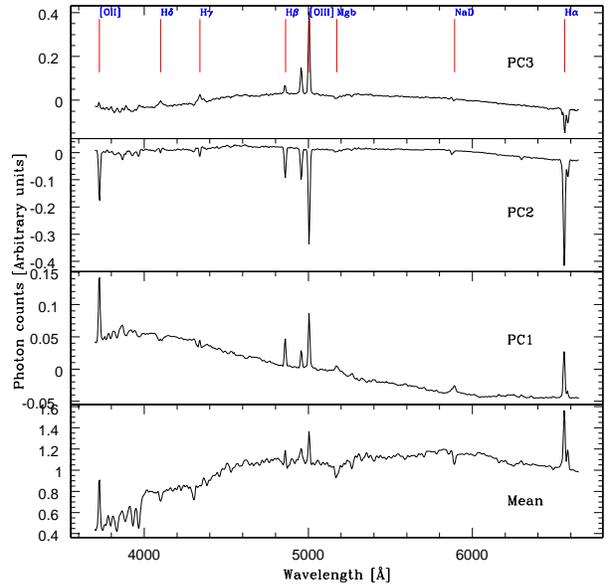,width=\plotwidth}}
\caption{The mean spectrum and first three principal components for the
sample spectra. Note that the sign of the PCs is arbitrary.}
\label{pcs}
\end{figure}

Principal Component Analysis of the sample spectra was carried out by
finding the eigenvectors (principal components) of the covariance
matrix of the de-redshifted and mean subtracted spectra. The mean
spectrum and first three principal components (PCs) can be seen in
Figure~\ref{pcs}. The first PC accounts for 49.6\% of the variance in
the sample, the second accounts for 11.6\%, and the third accounts for
4.6\%. This still leaves 34.2\% of the variance for the later
PCs. Much of this remaining variance will be due to noise in the
data---compare the distribution of variance over the principal
components with that seen in the PCA of synthetic spectra by Ronen
\etal\ (1999) or high S/N observations by Folkes (1998). The 1st PC
shows the correlation between a blue continuum slope and strong
emission-line features. The 2nd PC allows for stronger emission lines
without a strong continuum shape. The 3rd PC allows for an
anti-correlation between the oxygen and H$\alpha$ lines, relating to
the ionization level of the emission-line regions.

Figure~\ref{pc12classz} shows the distribution of the sample spectra
in the PC1--PC2 plane. The spectra form a single cluster, with the
blue objects with emission lines found to the right of the plot and
the red objects with absorption lines to the left. Objects with
particularly strong emission lines are found lower on the plot. We
expect that there exists some small scatter due to Poisson noise, but
this is suppressed due to the noise reduction property of the PCA
technique.  Errors in the flux calibration will cause a systematic
error, while fibre-to-fibre variations and changes from run
to run could introduce additional random scatter (see Ronen \etal\
(1999) and Folkes (1998) for a discussion of errors due to Poisson
noise and flux calibration).  For this reason, Bromley \etal\ (1998)
chose to high-pass filter the spectra. However this involves loss of
information, which we regard as undesirable.
The accuracy of the 2dF flux calibration is currently being
examined, and will be included in the final analysis of 2dF data.  

We will use the location of spectra in the PC1--PC2 
plane as the basis of our spectral classification scheme.
We have investigated the variation in the distribution of objects in
the PC1--PC2 plane as a function of various parameters (see Folkes,
1998). We find that there is very little difference in the
distribution with galaxy size or ellipticity. There is a small
variation with redshift, in that there is a population of low
luminosity galaxies with very strong emission lines at low redshift
which are not seen at higher redshifts. 

\section{SPECTRAL CLASSIFICATION}
\label{sec:class}

Principal Component Analysis has revealed the main features of the
galaxy spectra, but without some further information it is not clear
how to segment the PC1--PC2 plane into spectral classes, whether such
a classification is meaningful, and what the physical significance of
such classes would be. Two approaches were used in combination to gain
insight into the distribution of the galaxy spectra in principal
component space. One was to classify a subsample of the spectra by eye
using a simple phenomenological scheme, and hence look at the
distribution of galaxies with specific spectral features in the PC
space. The second was to take the Kennicutt (1992) sample of spectra
belonging to galaxies of known structural morphology and project them
onto the PC1--PC2 plane.
A third approach , to relate the PCs to physical parameters 
(e.g. 
age, metallicity and star-formation history of galaxies) 
by using model spectra, 
is discussed in
Ronen \etal\ (1999).

\subsection{Spectral features}
\label{ssec:features}

A simple 3-parameter classification scheme was used to identify
spectral features. The scheme allocates each galaxy spectrum a code
number from 0 to 2 according to the strength of spectral features
in each of the following three categories: 
early-type absorption lines (molecular features such
as H, K, CN, Mg), Balmer series absorption lines (H$\gamma$, H$\delta$, etc.),
and nebular emission lines (OII, OIII, H$\beta$ etc.). This is physically
motivated by the typical features produced by stellar populations
at progressive stages during and after star-formation.
We have selected an unbiased subsample of 56 galaxies which were 
classified using this scheme,
and then collected into  six broad classes with physical motivation, 
as follows: Class A, strong absorption lines; Class B, weak absorption lines;
Class C, weak features; Class D, strong balmer lines; Class E, strong
emission lines; Class F, Strong Balmer and Emission lines. 
We note that these classes are not directly related to any structural
morphology. 

\begin{figure*}
\centerline{\psfig{figure=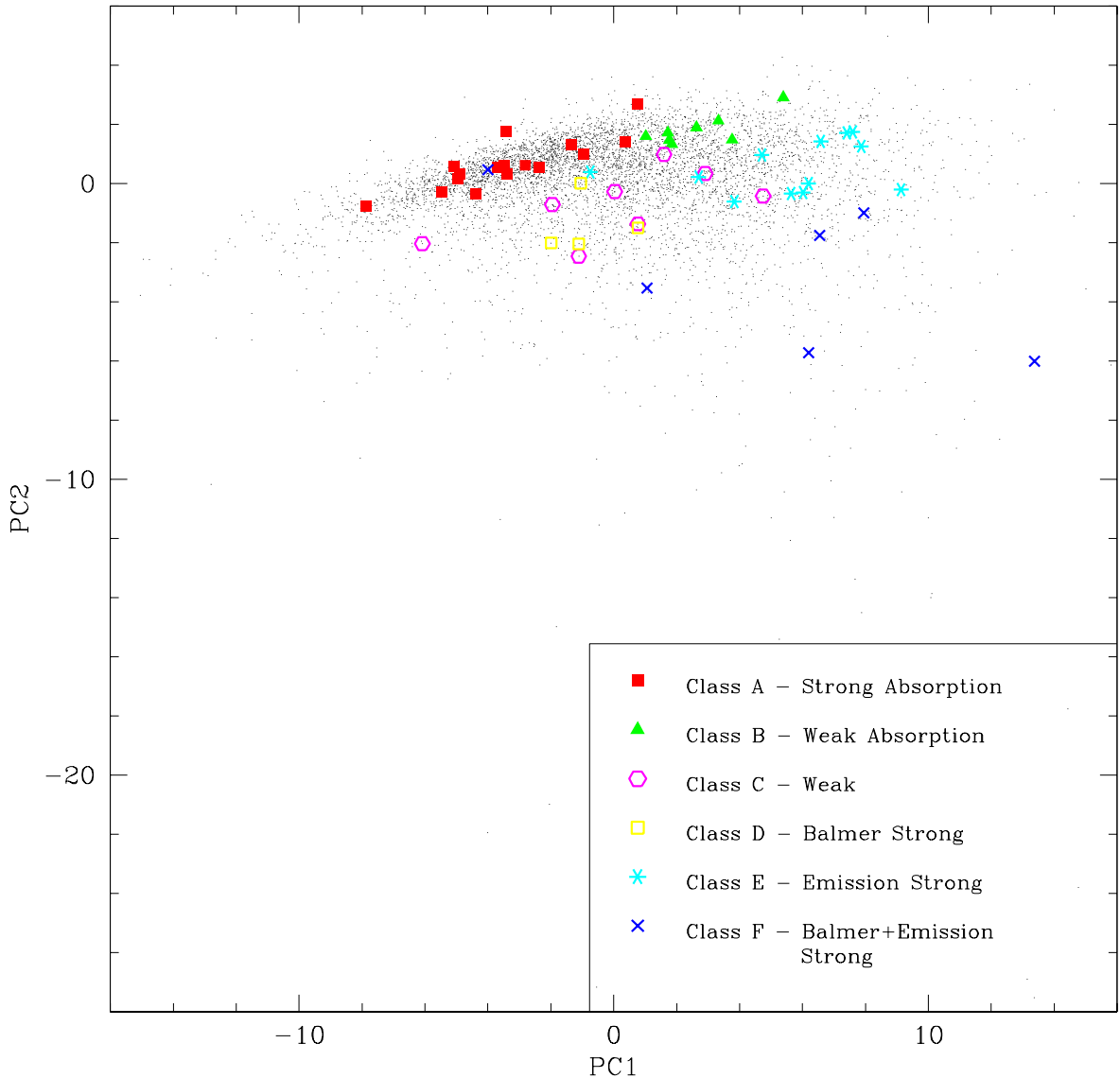,width=2.0\plotwidth}}
\caption{The distribution of galaxy spectra in the PC1--PC2 plane, with
the visual classifications based on spectral features overlaid for a
subset of the objects. }
\label{pc12classz}
\end{figure*}

The 56 example objects can be plotted on the PC1--PC2 plane. 
These can be
seen in Figure~\ref{pc12classz}, which shows considerable segregation on
the PC1--PC2 plane. The Class A objects, representing the strong
absorption line systems with old stellar content and little star
formation inhabit a clear region of the plots. The Class B, weaker
absorption line systems also show a clear cluster. Classes D, E and F
with emission and/or Balmer lines inhabit the lower sections of the plot
in fairly distinct areas. The Class C objects that do not have
particularly prominent features in absorption or emission are more
widely spread.

\begin{figure*}
\centerline{\psfig{figure=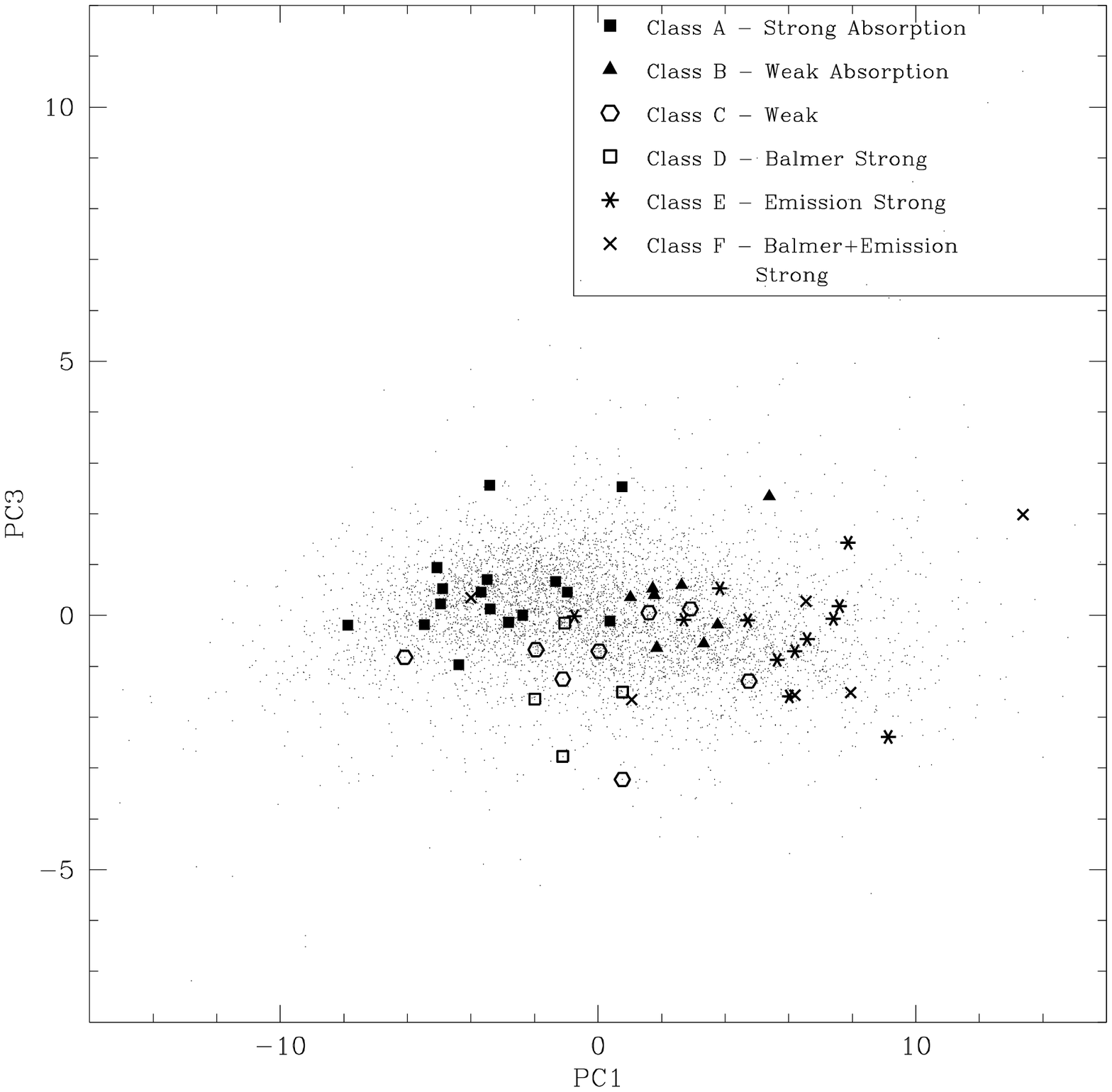,width=2.0\plotwidth}}
\caption{The distribution of galaxy spectra in the PC1--PC3 plane, with
the visual classifications based on spectral features overlaid for a
subset of the objects.}
\label{pc13class}
\end{figure*}

Although some segregation is shown in the PC1--PC3 plane, in general PC3
is not such a good discriminator, as shown in Figure~\ref{pc13class}.
However it does allow good separation of the Balmer and oxygen emission
line objects, since PC3 allows for an anti-correlation between those
lines.

\subsection{Galaxy morphology}
\label{ssec:morph}

In the second approach, the 55 Kennicutt (1992) galaxies were split into
five standard morphological groups (E/S0, Sa, Sb, Scd, Irr) plus 29
objects with unusual spectra. To make use of this set of well-fluxed,
reliable spectra of known morphology, they need to be projected onto the
PC space defined by the 2dF spectra. To do this, each Kennicutt galaxy
is de-redshifted to its rest frame then smoothed with a 3\AA\ Gaussian
filter, which (by experimentation) gives a line profile similar to that
of the 2dF spectra. The Kennicutt spectra are then sampled with 4\AA\
bins across the same wavelength range and normalised in the same way as
the 2dF data. The remaining uncertainties are due to any systematic
errors in the 2dF response function.

\begin{figure*}
\centerline{\psfig{figure=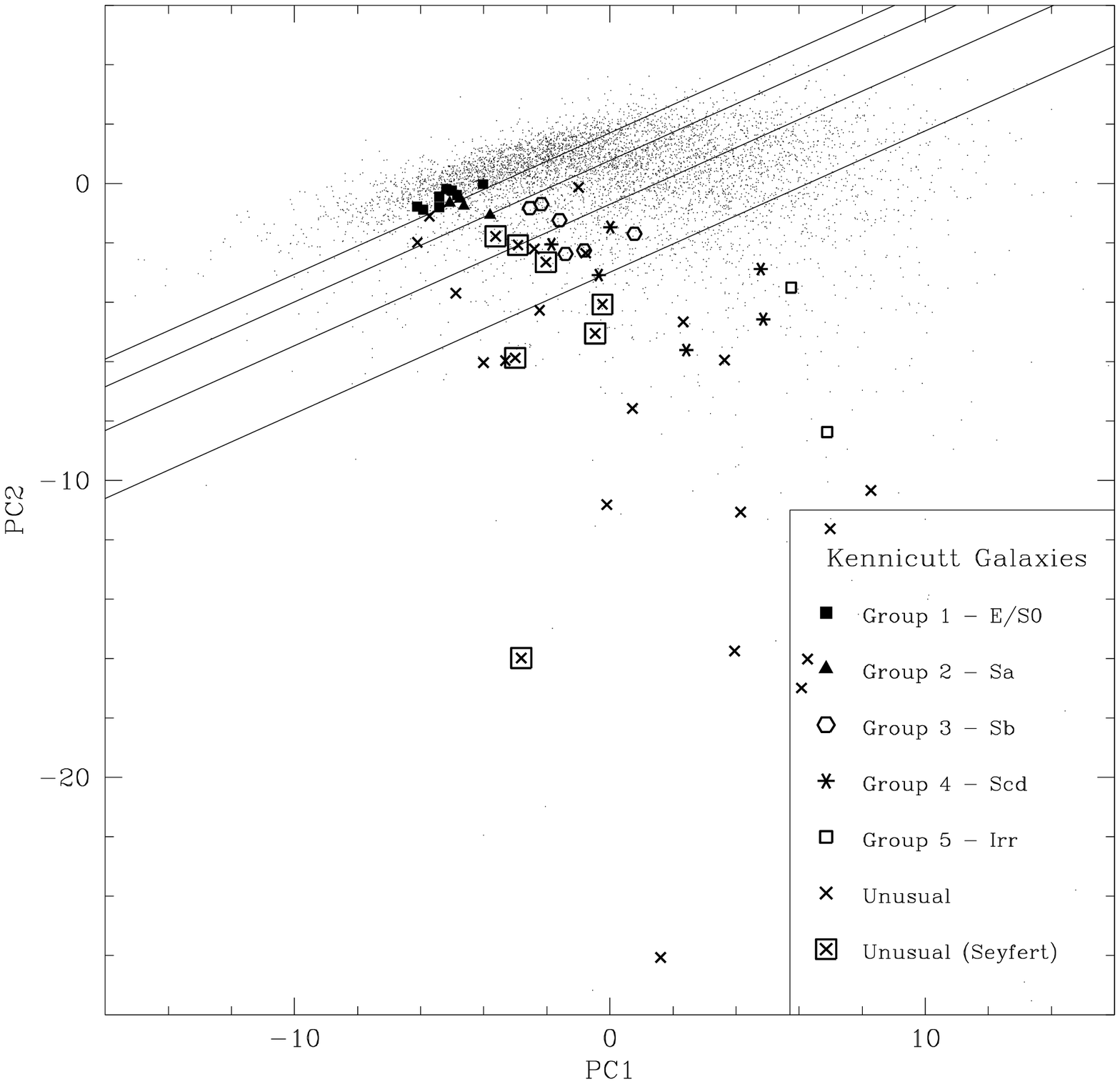,width=2.0\plotwidth}}
\caption{The distribution of the sample spectra in the PC1--PC2 plane.
The positions of the Kennicutt sample galaxies are overlaid. The lines
divide the plane into the five spectral classes we have adopted, with
type~1 objects towards the top left of the plane and type~5 objects
towards the bottom right.}
\label{cuts}
\end{figure*}

The 55 Kennicutt galaxies prepared in this way were then projected
onto the PCs from the 2dF sample. Figure \ref{cuts} shows the PC1--PC2
plane with the Kennicutt points labeled by morphological group. This
figure clearly shows the progression in morphological type across the
plot, with the many unusual objects, such as star-bursts and irregulars
populating the extreme emission-line regions. The Seyfert galaxies
from the Kennicutt sample fall below the morphological sequence of
normal galaxies, since they have emission lines that are not
necessarily associated with a blue continuum. However many of the
other unusual galaxies with star-burst activity also fall in this area,
so that the Seyferts are not clearly
segregated. Figure~\ref{pc12classz} shows that this area is populated
by the Balmer strong objects and some of the Class~3 survey spectra,
which show a variety of weak absorption and emission features.

\subsection{Definition of spectral types}
\label{ssec:types}

It is now possible to define sensible classifications in PC space
based on meaningful spectral and morphological classifications. Here
we wish to emphasise the links between galaxy spectra and morphology,
so we choose to employ parallel cuts in the PC1--PC2 plane along the
Hubble sequence as delineated by the Kennicutt galaxies. The cuts can
be seen in Figure~\ref{cuts} with the Kennicutt galaxies
superimposed. This defines five spectral types, which are roughly
analogous to the five morphological groups. 
The Kennicutt galaxies do not seem to fall in the region where
most of the 2dF galaxies are, and this may be due to the selection
bias of the sample, but also due to flux calibration.
With this reservation, the exact placement of the
lines on Figure~\ref{cuts} is somewhat arbitrary, but we have used
both the projection of the Kennicutt galaxies and our by-eye spectral
classification to place the lines as appropriately as possible. 
Note,  however,   that there is no one-to-one  
correspondence between these five
classes and the six classes described in section~\ref{ssec:features}.
To check the robustness of our object classification in view of the fact
that the response function is fibre dependent, we have examined 212
objects with repeated observations in overlapping fields. We find that 64\% of
the repeated objects have the same class, and 95\% have the same class to within one type.

\begin{figure}
\centerline{\psfig{figure=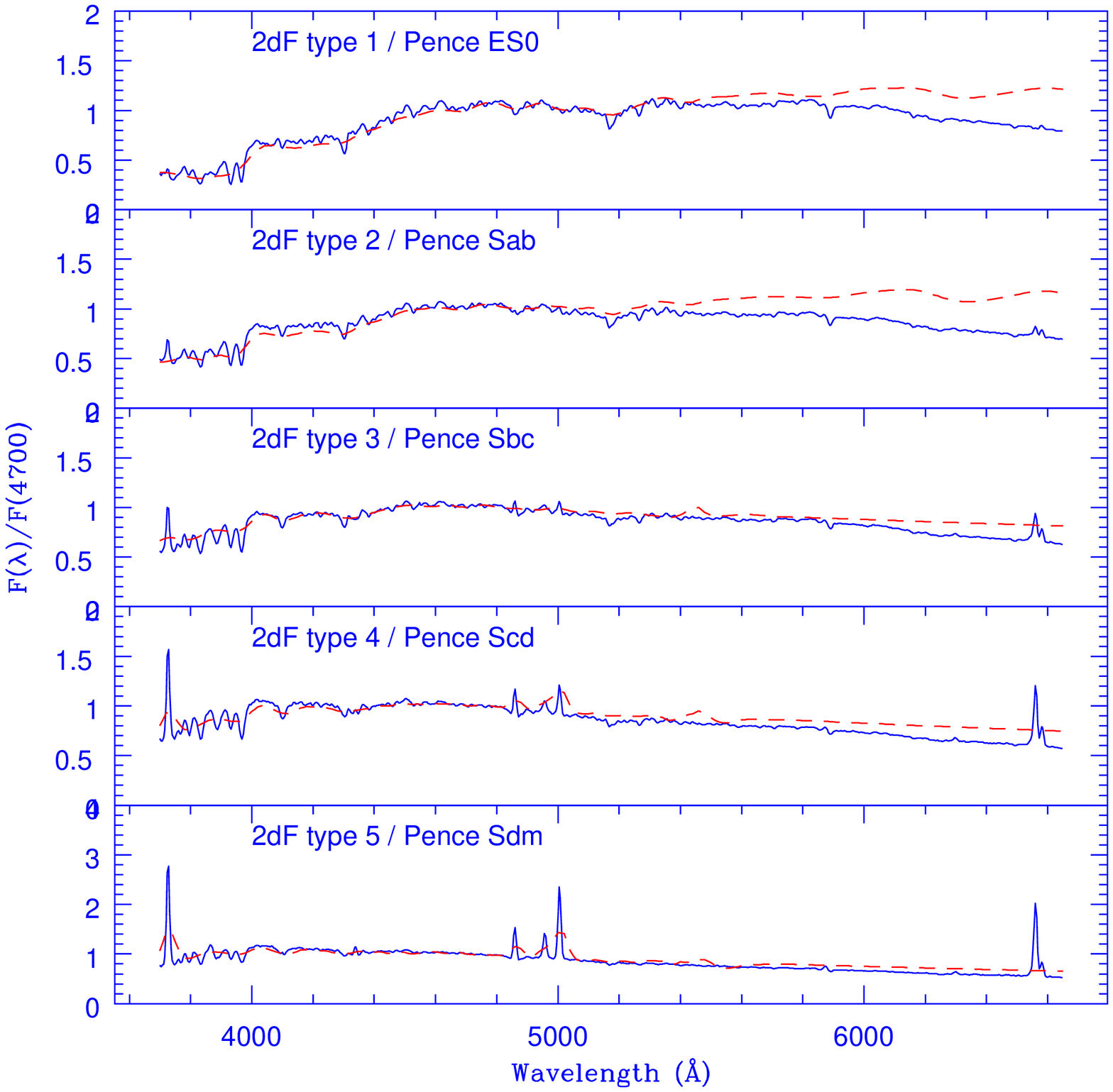,width=\plotwidth}}
\caption{The mean spectrum for each spectral type (solid line), compared 
with the corresponding Pence (1976) type (dashed line). }
\label{types}
\end{figure}

The mean spectra for each of the five types defined by these cuts are
shown in Figure~\ref{types}. This shows the clear progression from the
red absorption-line spectrum of Type~1 to the strong emission-line
spectrum of Type~5.  As can be seen from Figure~\ref{cuts}, there is
not a one-to-one relation between morphology and spectral type, but
the general relation is clear.  The mean spectra are in good agreement
with the spectra of the equivalent morphological groups given by Pence
(1976), Coleman \etal\ (1980) or Kennicutt (1992): Type~1 corresponds
approximately to E/S0 galaxies, Type~2 to Sa galaxies, Type~3 to Sb
galaxies, Type~4 to Scd galaxies and Type~5 to the irregulars. The
agreement is excellent in the blue end ($\lambda<5000\AA$), though towards the red end the
spectra of the 2dF types have less flux than the corresponding Pence
spectra. This is probably  due to inaccuracies in the current preliminary
flux calibration, and we will be making further observations to
improve the mean calibration of the 2dF spectra.
We note that, in principle, the PCs can be used as continuous variables, 
without binning them.

\subsection{K-corrections}
\label{ssec:kcorr}

Spectral types are of intrinsic interest, but are also important in
that they yield the K-corrections necessary for estimating absolute
magnitudes. The K-correction appropriate to a particular galaxy could
in principle be obtained directly from the observed spectrum, from its
PCA reconstruction, or from its principal components as $K(z,{\rm
PC1},{\rm PC2},\ldots)$. These approaches require careful examination
of a number of issues including the extent to which the fibre spectrum
is representative of the integrated galaxy spectrum, the systematic
uncertainties in the flux calibration and the available wavelength
range (cf Heyl et al, 1997).  A further complication with our current analysis is that we
have limited the PCA to a fixed range in rest-frame wavelength,
3700\AA\ to 6650\AA.  Since the $b_J$ pass-band extends from $3950$\AA 
to $5600$\AA, the wavelength range of the PCA reconstruction, or the
class mean spectra allow us to calculate K-corrections in $b_J$ only
for galaxies $z <0.07$.

\begin{figure}  
\centerline{\psfig{figure=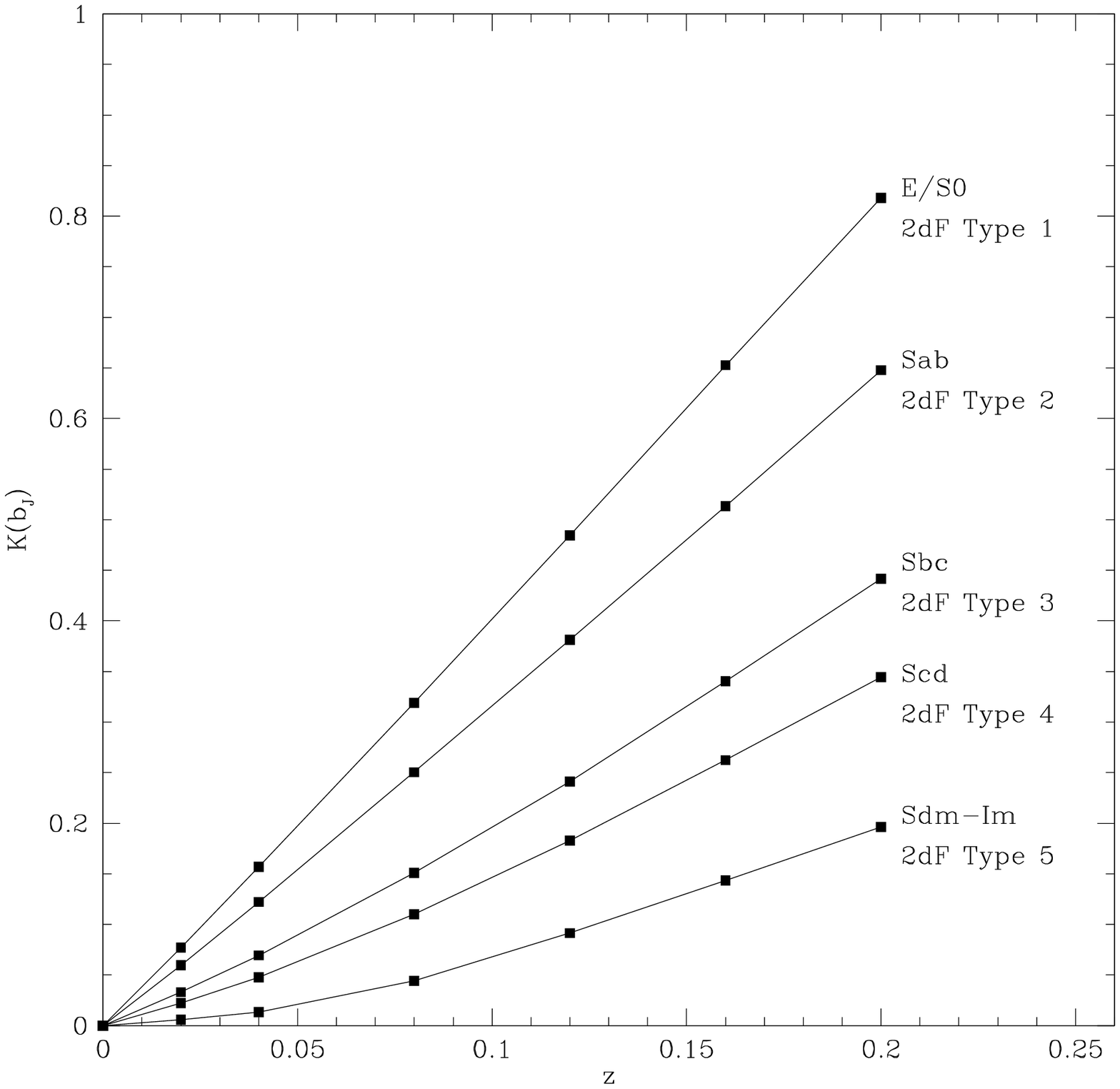,width=\plotwidth}}
\caption{K-corrections for the $b_J$ filter derived from values given in
Pence (1976) for B and V filter K-corrections.}
\label{kcorr}
\end{figure}

In light of these complications we adopt here a practical approach,
associating the spectral types defined in the previous section with
specific template spectral energy distributions (SEDs). As discussed
above, the SEDs of the five morphological types given by Pence (1976)
agree well (i.e, at the blue end, which is that relevant for the $b_J$
filter) with the mean spectra of our five spectral types. We therefore
identify our spectral types with Pence's SEDs. We compute the
appropriate K-corrections from Pence's tabulations, transforming the
K-corrections in the B and V filters according to the colour relation
given by Blair \& Gilmore (1982) for the $b_J$ filter,
\begin{equation}
K(b_J) = K(B) - 0.28 ( K(B) - K(V) ) \ .
\end{equation}
The K-corrections derived in this way for the redshift range $0<z<0.2$
are shown as the curves in Figure~\ref{kcorr}.

\section{THE GALAXY LUMINOSITY FUNCTION}
\label{sec:lfs}

\subsection{Methods}
\label{ssec:methods}

\begin{figure*}
\centerline{\psfig{figure=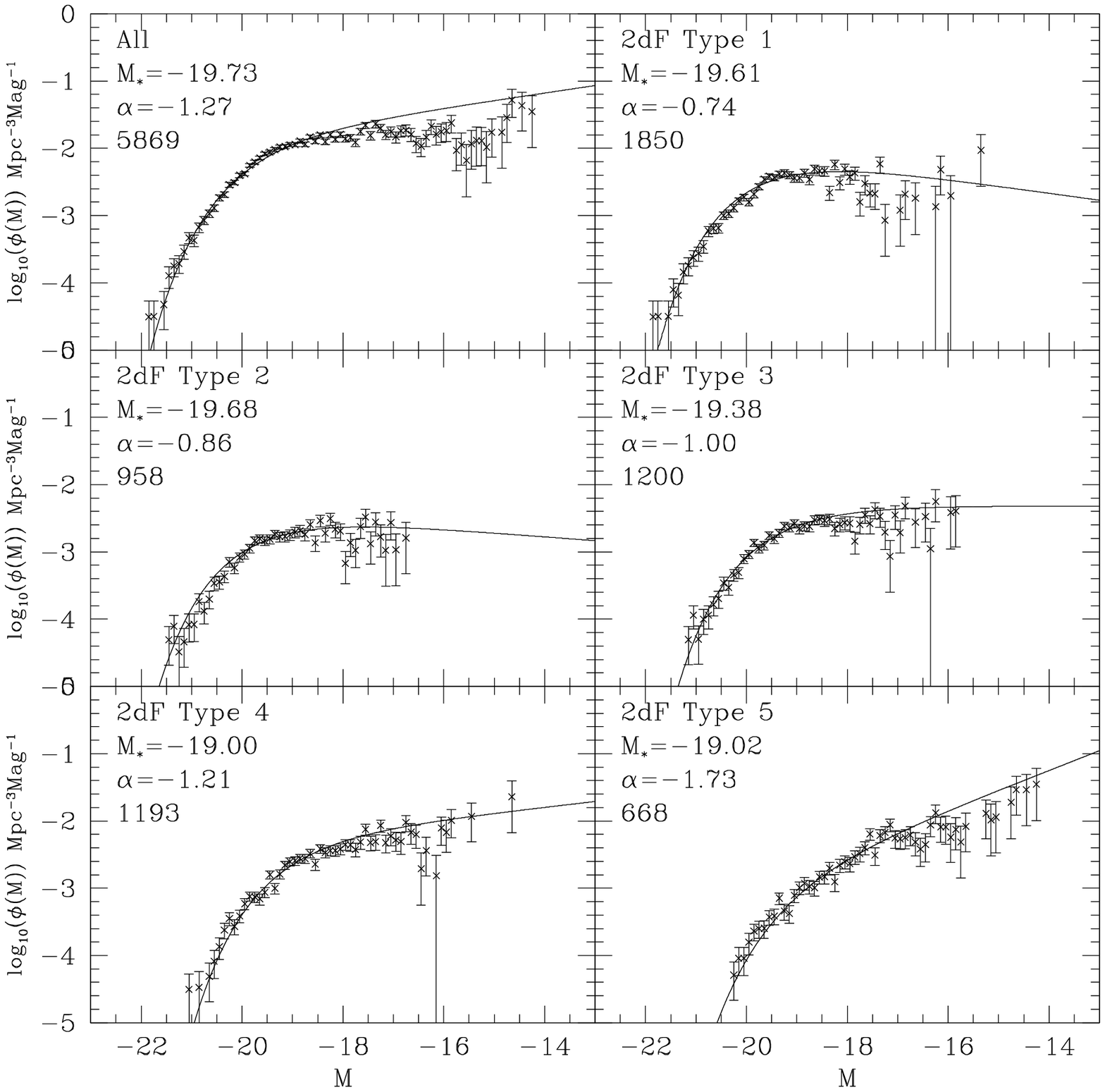,width=2.0\plotwidth}}
\caption{Luminosity functions including K-corrections and weighting to
compensate for incompleteness. Points show 1/$V_{\rm max}$ estimates
with Poisson error bars. 
This method assumes a uniform distribution, which is unlikely 
to be the case in our rather small survey volume.
The line is a Schechter function fit by the parametric
STY maximum likelihood method. The normalisations for the two methods
are derived separately (see text). The number of galaxies of each type
is indicated in the fourth line in each box. We used 
$\Omega$=1 and $H_0$=100\,km\,s$^{-1}$\,Mpc$^{-1}$.
}
\label{klum}
\end{figure*}

In computing the luminosity functions (LFs) we use both the 1/$V_{\rm
max}$ method (Schmidt 1968) for a non-parametric estimate of the LF
and the STY method (Sandage, Tammann \& Yahil 1979) for a maximum
likelihood fit of LF parameters. The statistical properties of these
different estimators are discussed by Felten (1976), Efstathiou, Ellis
\& Peterson (1988) and Willmer (1997). The 1/$V_{\rm max}$ method
assumes uniform spatial distribution, while the STY method assumes
a parametric form, here taken to be a Schechter function (Schechter 1976). The
use of other LF estimators is left for future work. We assume the
cosmological parameters to be $\Omega=1$ and $\Lambda=0$ (hence
$q_0=\frac{1}{2}$) and $H_0$=100\,km\,s$^{-1}$\,Mpc$^{-1}$.

The analysis is limited to $z<0.2$ by the definition of the sample
used in the PCA. The effective area of the survey can be estimated by
dividing the number of objects currently observed by the final
expected density. The density of galaxies brighter than $b_J=19.45$
as derived from the parent 2dFGRS source catalogue  is
$180/\sqdeg$. Taking into account a configuration completeness of
93\%, the effective area for the 7972 galaxies observed is $47.3
\sqdeg$. This provides the overall normalisation for our LF estimates.

We also define a completeness factor for each apparent magnitude range
in the currently observed sample compared to the parent photometric
sample, and weight each galaxy accordingly in our LF estimates.
We note that there may be selection biases 
that may depend on spectral type
(e.g, because of surface brightness effects and also 
because at low S/N emission line galaxies are easier to 
measure redshfits for). 
However it is not possible to account for such spectral type dependency 
in the completeness since, by defintion, the partition into classes is 
known only for our selected sample classified spectra and not for the 
parent catalogue.

Error estimates for the 1/$V_{\rm max}$ LFs are computed assuming
Poisson statistics to deduce the fractional error without weights, then
applying that fractional error to the actual estimate of the LF. These
errors are under-estimates since they neglect the effects of clustering,
which will be especially apparent at the faint end of the LFs, where the
galaxies are sampled over a relatively small volume.

Error estimates for the parameters obtained by the STY method are
found using the $\chi^2$ contours of the likelihood ratio
distribution. Note that the fact that the sample is limited to $z<0.2$
is irrelevant when finding the maximum likelihood Schechter
parameterisation, since the maximum likelihood method is based on
conditional probability given a redshift. The normalisation of the LF
is not given directly by the STY method, but it can be found by
integrating the derived LF over the observed volume of the survey and
comparing this to the actual number of galaxies observed.

The errors in the measured magnitudes lead to a Malmquist-like bias
which can have a noticeable effect on the LF. One method of correction
(\eg\ Loveday 1992) is to maximise the likelihood in the STY method
for a luminosity function which, when convolved with a Gaussian of
dispersion equal to the magnitude error, will give the observed data.
Another source of bias is the fact that isophotal magnitudes are a
function of redshift, due to point spread function effects and
cosmological dimming. We have applied a correction to the APM
isophotal magnitude of each galaxy to approximate its total magnitude,
but subtle biases could remain due to the initial isophotal selection
and these may influence the shape and normalisation of LFs (Dalcanton
1998).

\subsection{Results}
\label{ssec:results}

\begin{figure}
\centerline{\psfig{figure=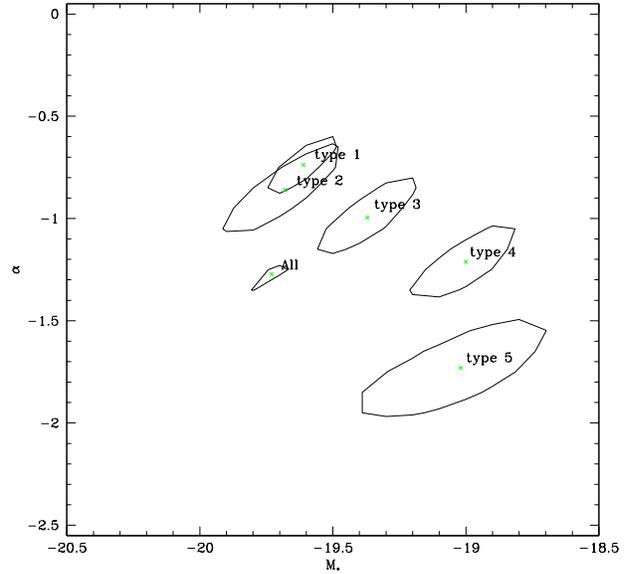,width=\plotwidth}}
\caption{Contours of likelihood for the Schechter function fits. The
contours show 95\% confidence levels. }
\label{kmaxall}
\end{figure}

Figure~\ref{klum} shows the 1/$V_{\rm max}$ LFs and the best-fitting
Schechter functions from the STY method for the whole sample and for
the individual spectral types. The parameters of the Schechter
functions are given in Table~\ref{tab:lffits}, while
Figure~\ref{kmaxall} shows the contours of likelihood for the
parameter estimates. Note that the number of galaxies in each
subsample has a large effect on the uncertainties. 
The errors on
M$^\star$ and $\alpha$ in Table~\ref{tab:lffits} define a box
which bounds a one sigma contour.
The LFs show a clear trend of fainter characteristic
magnitudes $M^\star$ and steeper faint-end slopes $\alpha$ going from
early to late spectral types. There seems to be a discrepancy between the 
1/$V_{\rm max}$ points and the STY curves at the faint end. This may indicate
that the LFs are not well fitted with a Schechter function, but may also arise
from the fact that there is  only a small number of faint galaxies and also
from the uniformity assumption of the 1/$V_{\rm max}$ method, which is not
assumed by the STY method.

\begin{table*}
\centering
\caption{Schechter function fits by spectral type. The errors given
for M$^\star$ and $\alpha$ define the smallest box that bounds a one
sigma contour. For $\phi^\star$, $\bar n$ and $\langle L \rangle $ one sigma errors are
given. Note: the galaxy density $\bar n$ is computed for galaxies brighter
than $M_{min}=-17$. $\langle L \rangle $ is the mean luminosity in units of
erg\,sec$^{-1}$\,Hz$^{-1}$\,Mpc$^{-3}$. 
We use $\Omega$=1 and $H_0$=100\,km\,s$^{-1}$\,Mpc$^{-1}$.
}
\label{tab:lffits}
\begin{tabular}{|c|r|r|r|r|r|r|} \hline
{Sample} & \multicolumn{1}{c}{\# gals} &\multicolumn{1}{c}{$M^\star$} & 
\multicolumn{1}{c}{\em $\alpha$} & \multicolumn{1}{c}{$\phi^{\star}\times
 10^{-3}$} & \multicolumn{1}{c}{$\bar n \times 10^{-3}$ Mpc$^{-3}$} &  
\multicolumn{1}{c} {$\langle L\rangle \times 10^{-25}$} \\ \hline
All    & 5869 & $-19.73\pm0.06$ & $-1.28\pm0.05$ & $16.9\pm1.7$ & $49.0\pm2.0$&$49.4\pm2.7$\\ 
Type 1 & 1850 & $-19.61\pm0.09$ & $-0.74\pm0.11$ & $9.0\pm0.9$ & $13.1\pm1.8$ &$29.3\pm3.0$ \\
Type 2 &  958 & $-19.68\pm0.14$ & $-0.86\pm0.15$ & $3.9\pm0.6$ & $6.5\pm1.2$ & $12.7\pm 2.0$\\ 
Type 3 & 1200 & $-19.38\pm0.12$ & $-0.99\pm0.13$ & $5.3\pm0.8$ & $9.1\pm1.2$ & $12.2\pm1.4$\\ 
Type 4 & 1193 & $-19.00\pm0.12$ & $-1.21\pm0.12$ & $6.5\pm1.3$ &
$11.1\pm1.1$& $9.9\pm1.1$\\ 
Type 5 &  668 & $-19.02\pm0.22$ & $-1.73\pm0.16$ & $2.1\pm1.1$ & $6.2\pm0.3$ &
$3.2\pm0.8$\\  \hline
\end{tabular}
\end{table*}

\begin{figure}
\centerline{\psfig{figure=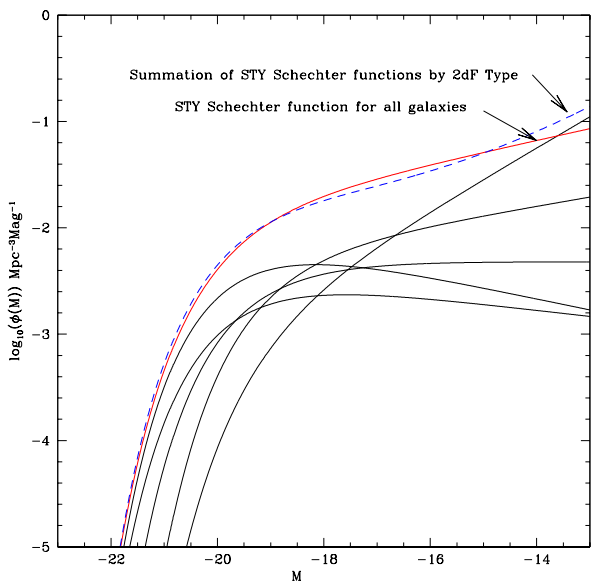,width=\plotwidth}}
\caption{The Schechter function fits for each spectral type are shown
based on the K-corrected and incompleteness-corrected data. The linescan be easily identified by the gradient of the faint-end slope which
steepens with spectral type (cf. Fig.~\ref{klum}). Also shown is the
Schechter function for all the galaxies, and a luminosity function formed 
by the summation of the individual Schechter
functions for the spectral types (dashed line).}
\label{klumadd}
\end{figure}

Note the peculiar result that $M^\star$ for the whole sample is brighter
than $M^\star$ for any of the individual spectral types. How this comes
about is illustrated in Figure~\ref{klumadd}, which shows the
co-addition of the luminosity functions for the individual spectral
types to give the total luminosity function,  and indicates the relative
contribution of the spectral types at each absolute magnitude. The most
remarkable point about this figure is the way that the very different
Schechter functions of the five spectral types combine to give an
overall LF that is also a Schechter function, at least down to
$M_{b_J}$=$-$16. Fainter than this the steep LF of the latest types
comes to dominate the overall LF, resulting in an upturn in the faint
end slope. The additional information provided by the spectral
classification is clear from this figure, and confirms the comment made
by Binggeli, Sandage \& Tammann (1988) that discussion of a luminosity
function without knowledge of the galaxy types is `covering a wealth of
details with a thick blanket'.

\begin{figure}
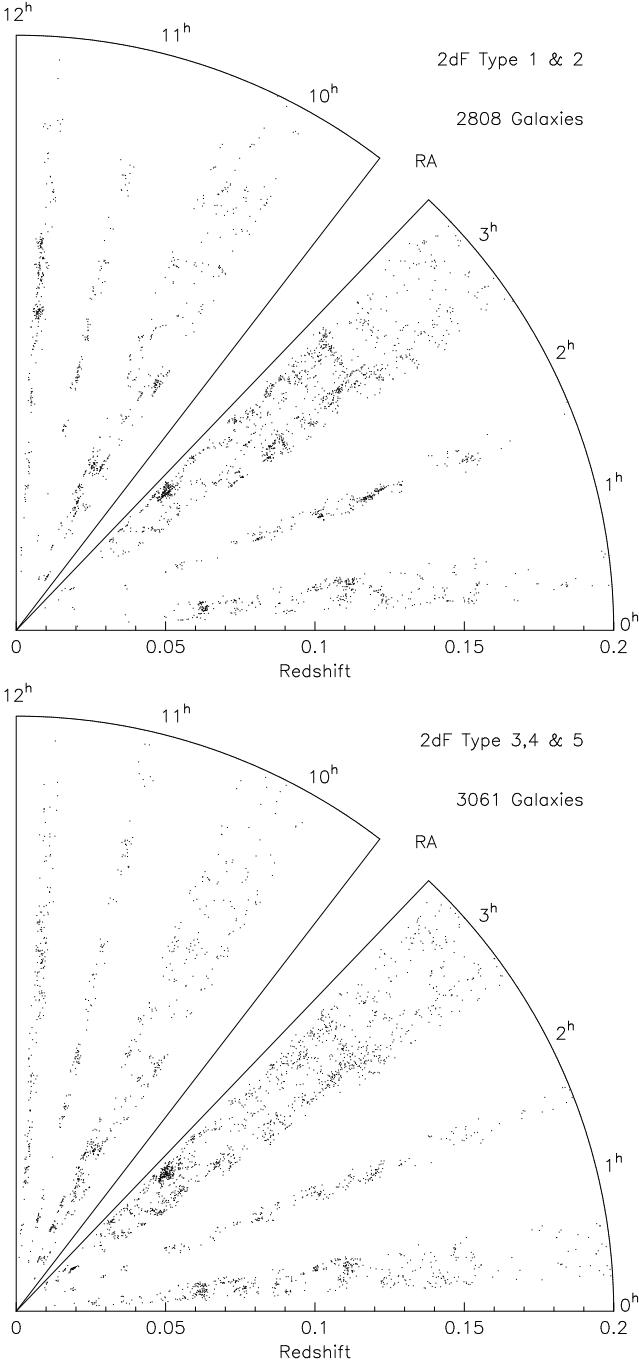

\centerline{\psfig{figure=SLF_n15a.ps,width=\plotwidth,angle=270}}
\centerline{\psfig{figure=SLF_n15b.ps,width=\plotwidth,angle=270}}
\caption{Redshift cone plots for the early-type (spectral types 1 \& 2)
and late-type (spectral types 3, 4 \& 5) galaxies in the sample. The
number of objects in each plot is sufficiently similar that a visual
comparison of the differences in redshift space clustering can be made.
The lower cone is for the south galactic pole strip and above it is a cone
for the north pole strip. }
\label{wedge}
\end{figure}

As well as looking at variations in the LFs with spectral type, we can
also provide a preliminary picture of the differences in clustering as
a function of spectral type. Figure~\ref{wedge} shows cone plots of
the redshift-space distribution of early types (types~1 and~2) and
late types (types~3, 4 and~5); these combinations were chosen simply
to give similar numbers of galaxies. The red, `early-type' galaxies do
appear more clustered, with evidence for `finger-of-God' effects
caused by the velocity dispersion of galaxy clusters, in agreement
with the long-known morphology-density relation (Dressler 1980). In
comparison, the blue, `late-type' galaxies show a more uniform
distribution, although clustering is still evident. Quantifying these
differences and comparing them with the predictions of models (\eg\
Cole \etal\ 1998) will be a major focus of future analysis of the
2dFGRS.

\section{DISCUSSION}
\label{sec:discuss}

\subsection{PCA and spectral types}
\label{ssec:pcatypes}

The 2dF system and the adopted survey strategy yield a good data set 
for spectral analysis. The broad wavelength coverage and the
homogeneity of the spectra are particularly important. The key
remaining unknown in the analysis is variation in the system response
as a function of time, fibre or location on the field plate. These
variations may influence the results of the PCA and produce some of
the scatter in the distribution of galaxies in PC space. Large adverse
effects are not apparent in the results presented here, but some of
the PCs beyond the third do show unphysical broad features which may
be due to variation in the response function. Since the PCs are an
orthogonal basis set, restricting the analysis to the first few PCs
means that these irregularities are not allowed to influence the
spectral classifications. The ideal, however, would be to obtain fully
fluxed spectra with the use of standard star observations for each
observing run, and rigorous testing of fibre throughput, and this may
become possible as the 2dF system continues to develop. Even without
this, it may still be possible with the full data set to determine
(and correct) the average strength of the PCs as a function of fibre
number or plate position. Further refinement of the procedures used to
remove the adverse effects of sky lines, atmospheric absorption and
bad pixels will also improve the quality and robustness of the
analysis.

The chosen rest-frame wavelength range is a major issue for the PCA
method. It would be possible to further restrict the wavelength range
so that the broader redshift range $0<z<0.3$ could be
included. However this would involve limiting the rest-frame spectra
to the wavelength range from 3650\AA\ to 6150\AA, excluding the
H$\alpha$ line from the analysis. An alternative method would be to
analyse the $0<z<0.2$ spectra and the $0.2<z<0.3$ spectra separately,
with the possibility of finding a relation between the classes found
in each case.

PCA is clearly extracting physical information from the spectra.
Figure~\ref{pcs} shows that the first PC emphasises the correlation
between the blue continuum and the emission line strength. The second
PC emphasises the emission lines alone, while the third PC allows for
an anti-correlation between H$\alpha$ and the [OIII] lines reflecting
different excitation levels. 
Ronen \etal\ (1999) show how PCA can be
used in conjunction with population synthesis or evolutionary
synthesis models of galaxy spectra to extract information regarding
the age, metallicity and star-formation history of galaxies.

Here we have simply used PCA for spectral classification. Our spectral
types were defined in the PC1--PC2 plane (Figure \ref{cuts}) by
reference to the locations of galaxies belonging to
morphologically-defined groups, with the intention of defining a
spectral sequence analogous to the Hubble sequence. There are a number
of refinements that can be considered here. In order to better define
the location and the spread of the morphology sequence 
on the PC1--PC2 plane, a larger set of
tracer objects is required than the 26 normal Kennicutt galaxies. For
this reason it would be very useful to observe high S/N integrated
spectra for a much larger set of galaxies covering the full range of
structural morphological types at a range of inclinations. Alternatively, some of the brighter
2dFGRS galaxies could be morphologically classified by eye or by
automated means (\eg\ Naim 1995; Abraham \etal\ 1994), or in some
cases using existing classifications from the partially overlapping
Stromlo-APM Redshift Survey (Loveday \etal\ 1992). This second
approach has the additional benefit of allowing studies of the links
between morphology and spectral type.

Another possible approach is to define a purely spectral classification,
along the lines presented in section 4.1. This has the
advantage of being self-consistent. A third possibility is to use
a training set from models (Ronen \etal\ 1998).

\subsection{Comparison with other luminosity functions}
\label{ssec:lfs}

It is useful to compare our results on the LF with other recent
measurements.  Ratcliffe \etal\ (1998) determine the LF for 2055
galaxies in the Durham/UKST Galaxy Redshift Survey and find
$M^\star(b_J)=-19.7$, $\alpha=-1.1$,
$\phi^\star=0.012$\,Mpc$^{-3}$. After correcting for Malmquist bias,
they find $M^\star(b_J)=-19.7$, $\alpha=-1.0$,
$\phi^\star=0.017$\,Mpc$^{-3}$; thus the correction for Malmquist bias
has the effect of dimming $M^\star$ and flattening the faint-end slope
a little which in turn raises the normalisation.  Zucca \etal\ (1997)
determine the LF for 3342 galaxies in the ESO Slice Survey and find
$M^\star(b_J)=-19.6$, $\alpha=-1.2$, $\phi^\star=0.020$\,Mpc$^{-3}$
after correcting for Malmquist bias.  Lin \etal\ 1996 use 18678
galaxies from the LCRS, and find $M^\star(R)=-20.$, $\alpha=-0.7$,
$\phi^\star=0.019$\,Mpc$^{-3}$.  Our present measurement of the
overall LF (see Table~2) is consistent  
with the ESO result (the largest pre-existing sample in $b_J$), 
but note that we have not yet corrected for Malmquist bias, 
and that this is expected to improve the agreement. We also note that all
the normalizations agree to within 10\%.  The SAPM
(Loveday \etal\ 1992) and SSRS2 (Marzke \& da Costa 1997) estimates of
$\phi^\star$ are $\sim 30\% $ lower than these estimates and probably
reflect a large local underdensity. We will carry out a more
detailed analysis of the overall LF and comparison to other surveys in
a future paper.

Loveday \etal\ (1992) determine LFs for the Stromlo-APM Redshift
Survey based on a sparse sample of galaxies to $b_J=17.15$. They
morphologically classify the images into E/S0, Sp/Irr and
unclassifiable samples.  They find Schechter parameters corrected for
Malmquist bias to be $M^\star(b_J)=-19.71\pm0.25$, $\alpha=0.2\pm0.35$
for the E/S0 sample and $M^\star(b_J)=-19.40\pm0.16$,
$\alpha=-0.8\pm0.20$ for the Sp/Irr sample. The steeply declining faint-end
slope for early type galaxies is due to the difficulty of classifying
faint, relatively featureless galaxies from the photographic images:
probably most of the un-classifiable galaxies are E/S0 galaxies. 
The SAPM luminosity functions based on emission line strengths
confirm this interpretation (Loveday et al 1998). 
So, the LFs for different morphological classes show the same 
trends as the LFs for the PCA classes considered here. 

Bromley \etal\ (1998) use a similar PCA analysis on spectra from the Las
Campanas Redshift Survey (Shectman \etal\ 1996). However their data
scaling and filtering means that the galaxy `clans' they derive are not
directly comparable to our spectral types. They do however confirm a
very similar progression in the faint-end slope of the Schechter
functions for the spectrally defined subsets, with values of $\alpha$
going from $\alpha=0.51\pm0.14$ for their earliest-type clan to
$\alpha=-1.93\pm0.13$ for their latest-type clan.

Lin \etal\ (1996) and Zucca \etal\ (1997)  split their samples
according to [OII] emission line equivalent widths. In both surveys
the strong emission line galaxies have a  faint-end slope that is
steeper by about 0.5, and $M^\star$ about 0.2 magnitudes fainter than
those without emission lines.  Loveday et al (1998) have also measured
the LF for samples split on H$\alpha$ and [OII] and find a similar
steep faint-end slope and fainter $M^\star$ for emission line
galaxies.  These variations are comparable to the changes in $\alpha$
and $M^*$ that we would find if we split our sample into just two PC
classes. 

This is an exploratory analysis and not the final word on the 2dFGRS galaxy luminosity
function. The complete survey sample will comprise about 40 times as many
galaxies, allowing investigation of the multivariate distribution of
galaxies over luminosity, spectral type, surface brightness and local
galaxy density. More sophisticated analyses will then be appropriate,
including: (i)~corrections for variations in completeness with redshift
and surface brightness as well as magnitude; (ii)~allowance for the
effects of clustering on the LF; (iii)~the use of clustering-independent
LF estimators; (iv)~correction for the Malmquist-like bias due to
magnitude errors; (v) tests of the physical significance and
robustness of the PCA spectral types, and (vi) aperture corrections. 

\section{CONCLUSIONS}
\label{sec:conclusion}

Spectral analysis and classification of 5869 2dF Galaxy Redshift Survey
spectra has been performed with a Principal Component Analysis 
method. The spectra form a sample limited to $b_J$=19.45 and
$0.01<z<0.2$. Methods have been applied to remove the effects of sky
lines, bad pixels, atmospheric absorption and the system response
function. The first PC was found to relate to the blue continuum and the
strength of the emission lines, while the second PC was found to relate
purely to emission line strength. The PC1--PC2 plane has been
investigated by classifying a subset of the spectra by eye on a
physically motivated spectral scheme, and also by projecting the
Kennicutt (1992) galaxy spectra of known morphology onto the plane. The
spectra have been classified into five spectral types with the mean
spectra of types~1 to~5 approximately corresponding to the spectra of
E/S0, Sa, Sb, Scd, and Irr galaxies respectively. Luminosity functions
for the spectral types have been computed, with type-specific
K-corrections and weighting of the galaxies to compensate for
magnitude-dependent incompleteness. Schechter fits to the luminosity
functions reveal a steadily steepening value of $\alpha$ and a trend
towards fainter $M^\star$ for later types. For spectral type~1 the
Schechter parameters are $M^\star=-19.61\pm0.09$ and $\alpha=-0.74\pm0.11$ 
whereas for spectral type~5 values of $M^\star=-19.02\pm0.22$ and
$\alpha=-1.73\pm0.16$ are found (errors define a box bounding the
one sigma contour). The
redshift-space distribution of spectral types~1 and~2 has been visually
compared to that of spectral types~3, 4 and~5, revealing qualitative
evidence for stronger clustering of the early-type galaxies. 
The methods used in this paper will form the basis of the analysis of the
luminosity function of the full 2dF Galaxy Redshift Survey.

\end{document}